# Self-Assembled Triangular Sc-rich Nanodomains: The Origin of Anomalous Alloy Scattering in ScAlN/GaN Heterostructures

**Authors:** Yuhao Yin,[1] Shizhao Fan[1,2,*]

[1] Division of Nano-Devices Research, Suzhou Institute of Nano-Tech and Nano-Bionics (SINANO), Chinese Academy of Sciences (CAS), Suzhou 215123, P. R. China

[2] Key laboratory of Semiconductor Display Materials and Chips, SINANO, CAS, Suzhou 215123, P. R. China

*Contact author: szfan2020@sinano.ac.cn

**Keywords:** ScAlN, alloy scattering, threading dislocation, strain field, two-dimensional electron gas

**Abstract**

Wurtzite scandium aluminium nitride (ScAlN) alloys hold substantial potential for GaN heterostructure high-electron mobility transistors, owing to their large spontaneous polarization gradient and the ability to achieve precise lattice matching at heterointerface. Nevertheless, despite continuous advances in epitaxy and device fabrication, the two-dimensional electron gas (2DEG) mobility in ScAlN/GaN heterostructures remains persistently limited, and the underlying microscopic origin remains elusive. Here, we uncover an anomalous electron scattering mechanism from laterally correlated alloy disorder that spontaneously develops within the ScAlN barrier driven by strain-field undulations imposed by threading dislocations. Dislocation strain field partitions the growth front into nanoscale domains with distinct strain states, driving anisotropic diffusion of Sc adatoms and yielding crystallographically oriented triangular Sc-rich nanodomains. Reducing the threading dislocation density from $2.8\times 10^{10}$ to $4.4\times10^{8}$ cm$^{-2}$ suppresses areal density of nanodomains by more than two orders of magnitude. As a result, low-temperature Hall mobility increases from 1800 to 11995 cm$^{2}\cdot$V$^{-1}\cdot$s$^{-1}$. These findings establish dislocation-mediated Sc-rich nanodomains as the dominant source of anomalously strong alloy scattering, resolving the mobility bottleneck for high-performance nitride electronics.

## Introduction

Achieving simultaneously high two-dimensional electron gas (2DEG) density and electron mobility in GaN-based heterostructures remains a fundamental materials challenge for high-power and high-frequency electronics.[1-4] Polarization engineering offers an effective strategy for increasing carrier density without intentional doping, yet conventional III-nitride barriers face inherent trade-offs among interfacial polarization gradient, strain accommodation, barrier thickness, and electron transport mobility.[3, 5] In AlGaN/GaN heterostructures, increasing the Al composition enhances the polarization discontinuity and boosts 2DEG density, but comes at the cost of elevated tensile strain, surface/interface roughness-induced scattering, and an increased Schottky barrier, which collectively degrade electron mobility and hinder Ohmic contact formation.[6-10] Utilizing a binary AlN barrier offers an even larger polarization discontinuity; however, its substantial lattice mismatch with GaN restricts the critical thickness to only a few nanometers. Ultrathin AlN barriers are susceptible to direct tunneling-induced gate leakage, whereas increasing the barrier thickness causes cracking and strain relaxation, compromising structural integrity and long-term device reliability.[11, 12] Alternatively, InAlN offers nearly lattice-matched growth on GaN at appropriate alloy compositions, enabling reduced strain while maintaining strong polarization. Nevertheless, its narrow epitaxial growth window, susceptibility to compositional fluctuations and phase separation, and limited thermal stability during high-temperature processing compromise heterointerface uniformity and device reliability.[13-17] These limitations have stimulated increasing interest in scandium aluminum nitride (ScAlN), an emerging high-κ ternary alloy with strong spontaneous polarization.[18-20] Incorporation of Sc into AlN allows composition-dependent lattice-matching with GaN, and the large spontaneous polarization discontinuity at ScAlN/GaN heterointerface enables exceptionally high 2DEG densities even with ultrathin barriers of 3-10 nm.[8, 21, 22] These distinctive attributes establish ScAlN/GaN heterostructures as a promising candidate to overcome the polarization–strain constraints of conventional III-nitride heterostructures and advancing the performance and reliability of HEMTs.

Recent advances in ScAlN/GaN heterostructures have demonstrated polarization-induced 2DEG densities exceeding $2 \times 10^{13}$ $cm^{-2}$ with ScAlN barrier thicknesses below 10 nm, enabling aggressive barrier scaling while maintaining strong electrostatic confinement.[18, 19, 23] This capability has facilitated substantial progress in device performance, including drain current densities reaching 4.0 A $mm^{-1}$ and high-power operation in the $K_a$-band frequency range.[24-27] Despite the exceptional 2DEG densities, the electron mobilities in ScAlN/GaN heterostructures remain considerably lower than those in optimized AlGaN/GaN counterparts, indicating strong scattering due to ScAlN barriers.[18, 28] Considerable efforts have therefore focused on improving epitaxial quality and interfacial abruptness. Streicher et al. achieved improved electron mobility by optimizing MOCVD growth conditions to suppress interfacial interdiffusion.[28] Wang et al. boosted electron mobility using a ScAlGaN quaternary alloy barrier, where Ga incorporation promotes surface adatom migration length, thereby improving surface smoothness and crystalline quality. [29] Casamento et al. further demonstrated that introducing an ultrathin AlN interlayer increased room-temperature mobility from approximately 300 to 1573 $cm^2/V\cdot s$, by suppressing electron wavefunction penetration into the alloy barrier and thereby mitigating alloy scattering. [18] Despite these advances, the root cause of the mobility bottleneck in ScAlN/GaN heterostructures remains poorly resolved.

In an ideal heterostructure, electron mobility is primarily governed by intrinsic phonon scattering, including polar optical phonon (POP), piezoelectric acoustic phonon (PZ), and acoustic deformation potential (ADP) scattering.[30, 31] POP scattering arises from the coupling between electrons and the macroscopic electric fields generated by longitudinal optical lattice vibrations in polar GaN, and becomes particularly important in the vicinity of room temperature.[31, 32] In contrast, PZ scattering originates from strain-induced polarization fields associated with acoustic phonons, whereas ADP scattering results from fluctuations in the conduction-band edge induced by lattice deformation.[32] However, practical ScAlN/GaN heterostructures are primarily affected by extrinsic scattering due to interface roughness (IR), threading

dislocation(DIS), and random alloy disorder. Interface roughness scattering arises from nanoscale fluctuations in the heterointerface profile, parameterized by the roughness amplitude ($\Delta$) and lateral correlation length ($\Lambda$). These fluctuations introduce local variations in the electron confinement potential for 2DEG.[31, 33] Advances in epitaxial growth have enabled root-mean-square (RMS) roughness down to approximately 0.5–0.7 nm for the surface of ScAlN/GaN heterostructures. Nevertheless, the resulting scattering strength also depends critically on the buried-interface morphology and lateral correlation length. [28, 34] Dislocation-related scattering arises from charged dislocation cores and their associated strain fields. Within established transport frameworks, however, this direct scattering channel remains relatively weak for threading dislocation densities (TDDs) of $10^8$-$10^9$ cm$^{-2}$, a regime representative of high-quality GaN heterostructures.[35] Alloy disorder (AD) scattering, in contrast, originates from local potential fluctuations associated with compositional disorder. Notably, Casamento et al. found that fitting the experimentally observed low mobility in ScAlN/GaN heterostructures with a conventional alloy-scattering model would require an anomalously large effective scattering potential of approximately 50 eV, far exceeding the 1.3–2.3 eV reported for AlGaN and InGaN alloys.[18, 30] However, the microscopic origin of this pronounced alloy scattering has not been identified.

In this work, we reveal the microscopic origin of alloy scattering in ScAlN/GaN heterostructures by combining temperature-dependent Hall measurements with multi-dimensional structural and transport characterization. Transport analysis identifies alloy disorder scattering as a dominant mobility-limiting mechanism, while reducing the threading dislocation density from $2.78 \times 10^{10}$ to $4.8 \times 10^8$ cm$^{-2}$ leads to a remarkable enhancement in low-temperature electron mobility from 1800 to 11995 cm$^2\cdot$V$^{-1}\cdot$s$^{-1}$, alongside a reduction of the effective scattering potential from 30 eV to 10 eV. Scanning transmission electron microscopy (STEM) and elemental mapping reveal spontaneously formed triangular Sc-rich domains within the ScAlN barrier. Further two-beam diffraction-contrast TEM and geometric phase analysis established a strong spatial correlation between these nano-domains and the anisotropic strain fields emanating from threading dislocations, indicating that dislocation-induced local strain promotes preferential Sc adatom agglomeration during epitaxy. These findings establish a direct link between dislocation-mediated alloy inhomogeneity and electron mobility degradation, offering new insights into the microscopic origin of alloy scattering and outlining a materials engineering route toward simultaneous high 2DEG density and high electron mobility in ScAlN/GaN heterostructures.

## Method

### Epitaxial Growth of ScAlN/GaN Heterostructures

ScAlN/GaN heterostructures were epitaxially grown on 4H-SiC substrates using a self-customed plasma-assisted molecular beam epitaxy (PA-MBE) system. Active nitrogen species were supplied by a radio-frequency nitrogen plasma source, while Ga, Al, and Sc were supplied from independently controlled thermal effusion cells. The substrate temperature was maintained at 750 °C throughout the epitaxial growth process. The group-III-to-active-nitrogen atomic flux ratio (III/N) was controlled to regulate the growth conditions of individual nitride layers. Specifically, an III/N ratio of 1.4 was employed for GaN and AlN growth under metal-rich conditions. The Sc content was controlled by adjusting the Sc/Al flux ratio. The barrier layer was grown under metal-rich conditions by supplying an additional Ga flux, while the (Sc + Al)/N flux ratio was maintained at 0.78.[36] The ScAlN/GaN heterostructures studied here employ a quaternary ScAlGaN barrier, following the community convention that groups Sc-incorporated AlN-based barriers under the ScAlN/GaN terminology. The heterostructure consisted of a 160 nm AlN buffer layer, a 640 nm GaN channel layer, a 2 nm AlN interlayer, a 1.5 nm GaN interlayer, a 10 nm Sc-containing barrier layer, and a 2 nm GaN cap layer, sequentially deposited on the 4H-SiC substrate. The ultrathin AlN and GaN interlayers were incorporated between the ScAlN barrier and the GaN channel to improve the heterointerface smoothness. ScAlN/GaN heterostructures with different TDDs were grown to investigate local Sc compositional inhomogeneity and electron transport properties.

### Structural and Microstructural Characterization

The surface morphology of the as-grown heterostructures was characterized using atomic force microscopy (AFM, Bruker, USA) and field-emission scanning electron microscopy (FE-SEM, Hitachi S-4800, Japan). The RMS surface roughness was extracted from AFM micrograph to evaluate the surface quality. The crystalline quality and structural properties were investigated using high-resolution X-ray diffraction (HRXRD, PANalytical X'Pert³ MRD, Malvern Panalytical) with Cu Kα radiation (λ = 1.54056 Å). Symmetric (002) and asymmetric (102) reflections were examined using 2θ–ω scans and ω rocking curves. The full widths at half-maximum (FWHMs) of rocking curves were analyzed to assess crystalline mosaicity and estimate TDDs.

TEM and STEM were performed using a Talos F200X microscope (Thermo Fisher Scientific) to investigate the microstructural features of ScAlN/GaN heterostructures. Two-beam diffraction-contrast TEM imaging was employed to visualize dislocation-related contrast and local strain variations, while STEM imaging was used to examine the morphology and spatial distribution of structural defects. Energy-dispersive X-ray spectroscopy (EDS) elemental mapping was conducted to resolve the distributions of Sc, Al, and Ga and identify local compositional inhomogeneities within the alloy barrier. Atomic-resolution STEM characterization was further performed using a probe aberration-corrected field-emission transmission electron microscope (Themis Z, Thermo Fisher Scientific) equipped with an EDS system. High-angle annular dark-field (HAADF) STEM imaging and elemental mapping were used to investigate the atomic-scale structure, local chemical variations, and their spatial correlations with threading dislocations. Geometric phase analysis (GPA) of the atomic-resolution STEM images was performed to evaluate local lattice distortion and strain distributions surrounding dislocation cores and Sc-rich domains.

**Temperature-Dependent Hall Measurements**

Temperature-dependent electrical transport properties were investigated using a cryogen-free Physical Property Measurement System (PPMS DynaCool, Quantum Design, USA). Temperature-dependent Hall measurements were performed from 2 to 300 K using the van der Pauw configuration. The 2DEG sheet carrier density and sheet resistance were extracted from the Hall and longitudinal resistance measurements, respectively, from which the Hall mobility was calculated. Temperature-dependent mobility was analyzed by considering multiple scattering mechanisms, including POP, PZ, ADP, IR, DIS, and AD scattering. The contributions of individual scattering mechanisms were evaluated within a semiclassical transport framework, and the total mobility was calculated using Matthiessen's rule. The extracted scattering parameters were correlated with the structural and compositional characteristics to identify the dominant mechanism governing electron transport in ScAlN/GaN heterostructures.

## Result and discussion

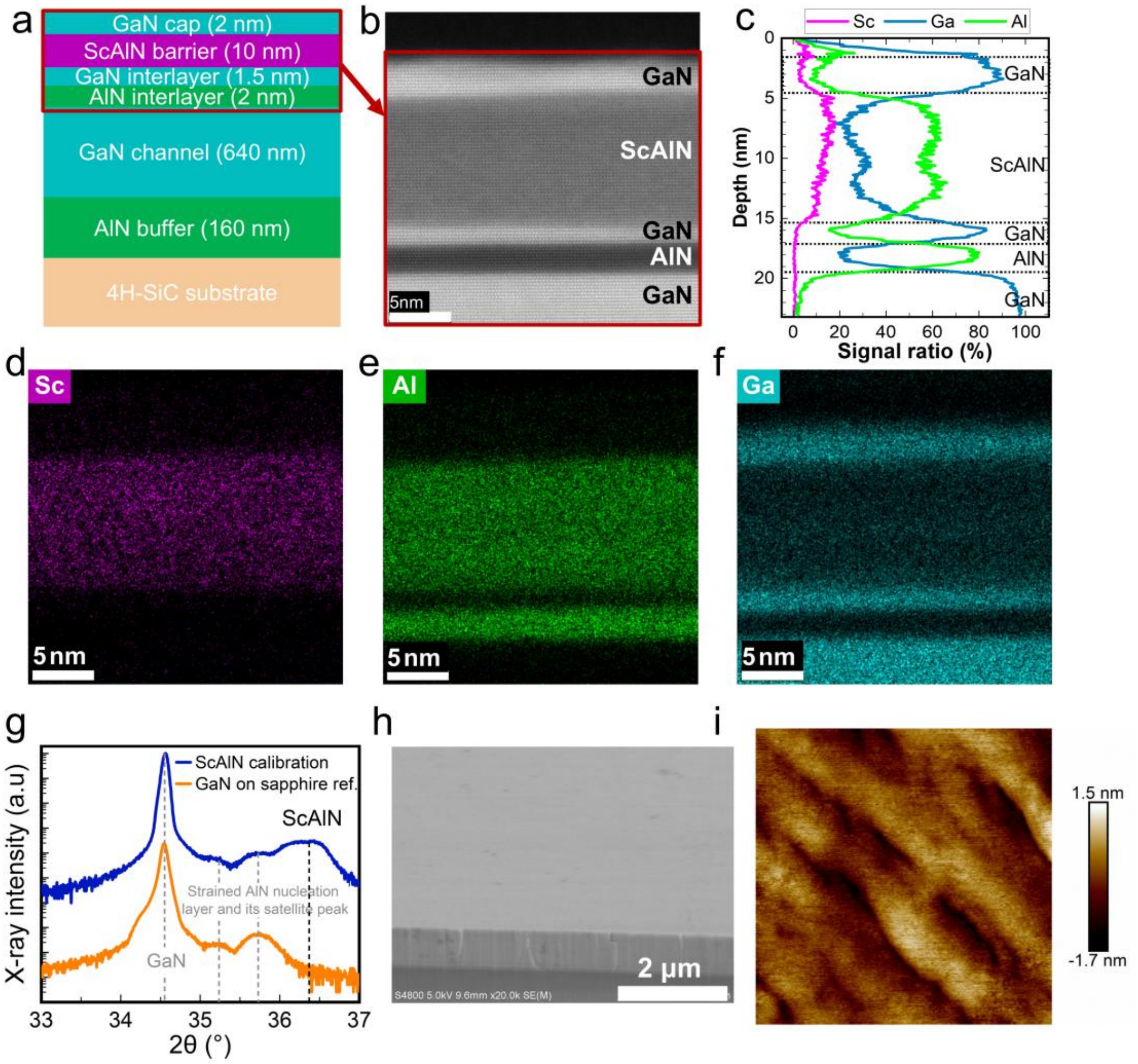

**Figure 1. Structural and compositional characterization of the ScAlN/GaN heterostructure.** (a) Schematic illustration of the epitaxial stack grown on a 4H-SiC substrate. (b) Cross-sectional STEM micrograph showing the multilayer architecture and atomically sharp heterointerfaces. (c) EDS line profiles tracking the depth-resolved elemental distributions of Sc, Al, and Ga along the growth direction. (d)–(f) Cross-sectional EDS elemental maps of d) Sc, e) Al, and f) Ga, respectively. (g) HRXRD 2θ–ω scans of a reference ScAlN calibration film and the underlying GaN template. (h) Cross-sectional SEM micrograph of the heterostructure. (i) AFM height map acquired over a 3 × 3 μm² scan window for the as-grown surface, with an extracted RMS roughness of 0.43 nm.

Cross-sectional STEM reveals a well-defined multilayer structure with sharp and coherent heterointerfaces, evidencing the successful epitaxial growth of the ScAlN/GaN heterostructure featuring a quaternary ScAlGaN barrier (Fig. 1b). The corresponding EDS depth profiles (Fig. 1c) show indicating negligible interfacial elemental interdiffusion. Along the growth direction, the relative atomic concentrations of Sc, Al, and Ga in the barrier layer fall within 14—16%, 64—58%, and 27—20%, respectively, consistent with the target design. The cross-sectional EDS elemental maps (Figs. 1d–f) corroborate these depth profiles and visually verify that Sc is spatially confined within the alloy barrier, from which an average barrier composition of $Sc_{0.14}Al_{0.64}Ga_{0.22}N$ is extracted. In Fig. 1g, HRXRD (002) 2θ–ω scans of the GaN-on-sapphire template and a 100-nm-thick ScAlN reference film are compared. A separate ScAlN (0002) reflection appears at ~36.3°, well resolved from the Bragg reflections of the GaN buffer and the 25-nm-thick strained AlN nucleation layer. The peak shift with respect to wurtzite AlN manifests Sc-mediated alteration of the out-of-plane lattice parameter $c$, whereas epitaxial elastic constraint locks the in-plane lattice parameter $a$ in close registry with GaN.[37, 38] The SEM image (Fig. 1h) reveals a macroscopically smooth surface without cracks or V-pits. The 3×3 $\mu m^2$ AFM micrograph (Fig. 1i) exhibits terrace-like surface undulations with a RMS roughness of only 0.43 nm, comparable to values reported for structurally optimized ScAlN/GaN heterostructures.[29] This exceptional surface smoothness provides a critical foundation for isolating the contribution of interface-roughness scattering in the subsequent transport analysis.

To investigate the electron transport properties of the as-grown ScAlN/GaN heterostructure, temperature-dependent Hall measurements were performed on Sample A from 10 to 298 K. Figures 2a–c present the temperature dependence of the Hall mobility, sheet resistance, and 2DEG density, respectively, with the corresponding electrical parameters summarized in Table I. As shown in Fig. 2a, the electron mobility increases from 922 $cm^2\ V^{-1}\ s^{-1}$ at 298 K to approximately 1800 $cm^2\ V^{-1}\ s^{-1}$ at 10 K. A plateau persists below ~100 K, followed by a precipitous fall at elevated temperatures. This behavior signals the onset of POP scattering above 100 K, while the low-temperature plateau is governed by extrinsic scattering. Correspondingly, the sheet resistance decreases from 283 to 143 $\Omega\ \square^{-1}$ upon cooling (Fig. 2b), whereas the 2DEG density remains nearly constant at approximately 2.4 × $10^{13}$ $cm^{-2}$ across the investigated temperature range (Fig. 2c). The negligible temperature dependence of carrier density confirms a robust polarization-induced electron channel, devoid of appreciable carrier freeze-out. Notably, the mobility rises by merely a factor of ~two despite strong phonon-scattering suppression at cryogenic temperatures, underscoring extrinsic scattering as the prevailing bottleneck for electron transport.

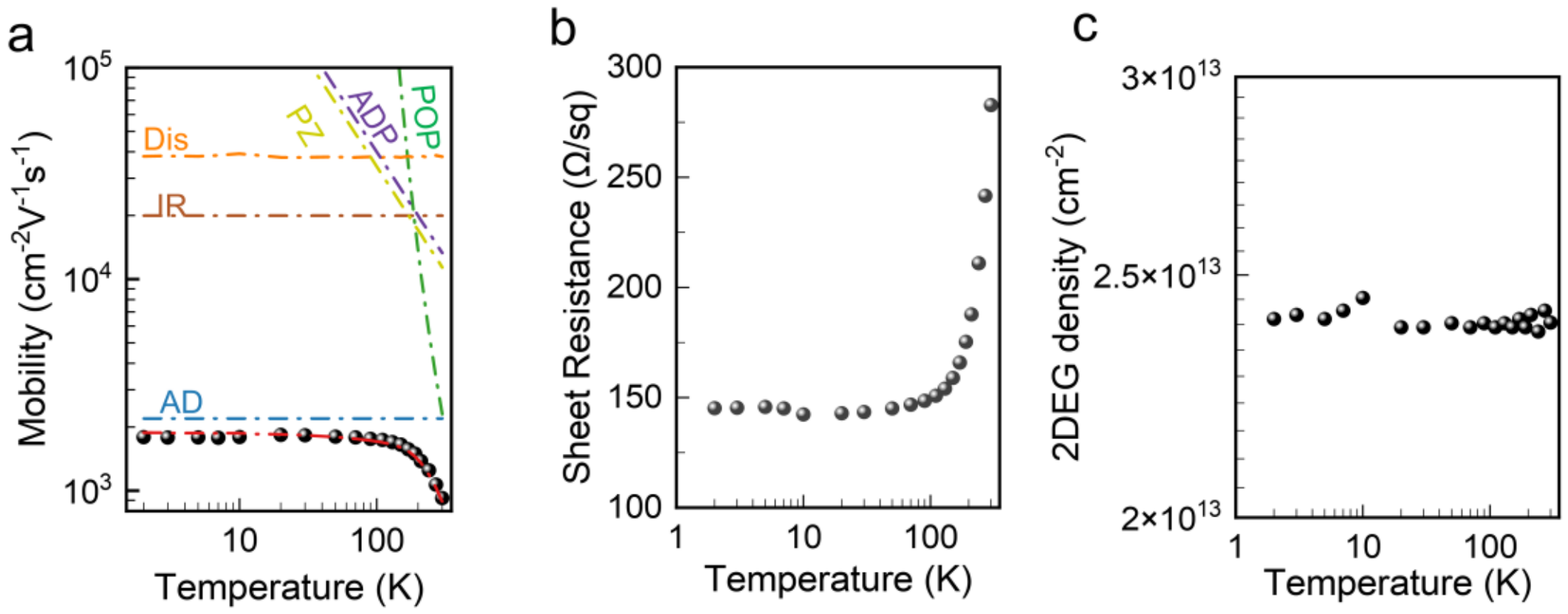

**Figure 2. Temperature-dependent electron transport and scattering analysis of Sample A.** (a) Experimental Hall mobility (symbols) and calculated mobility contributions from individual scattering channels, including alloy disorder (AD), interface roughness (IR), threading dislocations (DIS), acoustic deformation potential (ADP), piezoelectric acoustic phonons (PZ), and polar optical phonons (POP). The red dashed curve shows the total mobility calculated by Matthiessen's rule. Temperature-dependent (b) sheet resistance and (c) 2DEG sheet carrier density.

**Table I.** Summary of the temperature-dependent Hall transport data for Samples A and B measured at 298 and 10 K, including sheet resistance ($R_{sh}$), Hall mobility, and 2DEG sheet density.

| Sample | T(K) | $R_{sh}$ (Ω/□) | Mobility ($cm^2$/V·s) | 2DEG (/$cm^2$) |
|---|---|---|---|---|
| A | 298 | 283 | 922 | $2.39×10^{13}$ |
| | 10 | 143 | 1800 | $2.40×10^{13}$ |
| B | 298 | 176.6 | 1290 | $2.75×10^{13}$ |
| | 10 | 33.0 | 11995 | $1.60×10^{13}$ |

To quantitatively identify the dominant scattering mechanisms, the temperature-dependent Hall mobility was analyzed using a semiclassical transport model incorporating intrinsic phonon scattering and extrinsic disorder scattering, with the total mobility calculated according to Matthiessen's rule (see Section S2, Supporting Information). The intrinsic mechanisms include POP, PZ, and ADP scattering, which arise from electron interactions with optical and acoustic lattice vibrations in GaN. As shown in Fig. 2a, POP scattering grows in significance at elevated temperatures and drives the observed mobility drop, whereas PZ and ADP scattering yield comparatively weaker limitations across most of the investigated temperature range. To deconvolve the individual weight of different extrinsic scattering channels, we evaluated contributions from interface-roughness, threading-dislocation, and alloy-disorder scattering. Adopting an interface-roughness amplitude of 0.53 nm and a lateral correlation length of 9 nm, the calculated IR-limited mobility is approximately $2.0\times10^4$ $cm^2\cdot V^{-1}\cdot s^{-1}$, substantially exceeding the measured low-temperature mobility. Similarly, the direct DIS contribution, evaluated using a TDD of $2.78\times10^{10}$ $cm^{-2}$ (Fig. 4a), yields a considerably higher mobility limit than observed experimentally. These results indicate that neither interface roughness nor direct dislocation scattering alone can explain the substantial mobility degradation in Sample A. Instead, the experimental mobility is primarily constrained by alloy-disorder scattering, as evidenced by the close proximity of the dashed-blue curve (Fig. 2a) to the measured low-temperature plateau. To reproduce the experimental T-dependent mobility trend, an effective alloy scattering potential of approximately 30 eV is required, substantially exceeding the 1.3–2.3 eV values reported for conventional AlGaN and InGaN alloys.[18, 39, 40] This extraordinarily large effective potential points to strong local potential fluctuations tied to alloy inhomogeneity within the Sc-containing barrier. Nevertheless, transport fitting on its own cannot pinpoint the underlying microscopic structural origin, prompting direct examination of spatial Sc distribution and its correlation with threading dislocations and local strain fields.

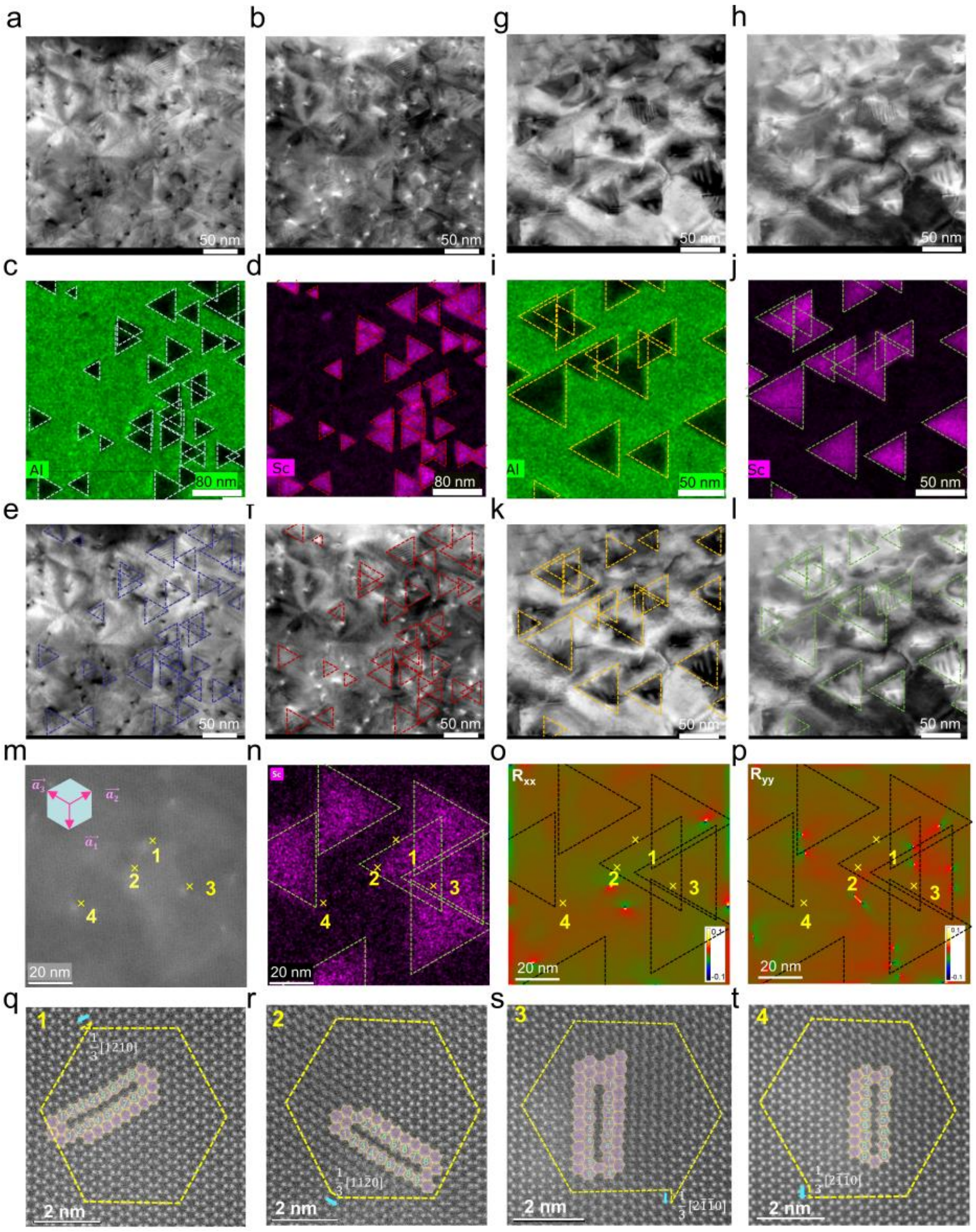

**Figure 3.** Atomic-scale structural characterization of Sample A. (a) Dark- and (b) bright-field TEM micrographs acquired under two-beam diffraction conditions. (c) Al and (d) Sc EDS maps recorded from the same region shown in (a) and (b), revealing triangular Sc-rich nanodomains. Extracted nanodomain boundaries are overlaid onto panels of (a) and (b) to yield (e) and (f). (g) Dark- and (h) bright-field micrographs acquired in STEM mode. (i) Al and (j) Sc EDS maps recorded from the same region shown in (g) and (h). Similarly, extracted nanodomain overlaid onto panels of (g) and (h) to yield (k) and (l). (m) A zoomed-in STEM micrograph highlighting a cluster of threading dislocations. Atomic-resolution micrographs of individual dislocations in panels (q)–(t) resolve distinct Burgers vectors. (n) Sc EDS map showing nanodomains bounded by dislocations at their peripheries. Geometric phase analysis (GPA) of panel (m) yields strain-field components along the $\langle 1\bar{1}00 \rangle$ ($R_{xx}$) and along $\langle 11\bar{2}0 \rangle$ ($R_{yy}$) crystallographic orientations, shown in (o) and (p), respectively.

To elucidate the origin of strong alloy disorder scattering, detailed microstructural investigations were performed on Sample A using TEM, STEM, and EDS. The STEM micrographs in Figs. 3a,b reveal numerous triangular nanodomains bearing distinct alternating bright and dark fringes, intertwined with threading dislocation outcrops. Notably, these triangular nanodomains exhibit well-defined boundaries aligned along the $\langle 1\bar{1}00 \rangle$ crystallographic directions and characteristic periodic fringes resembling moiré patterns, suggesting local variations in crystallographic orientation or lattice spacing.[41] The corresponding EDS elemental maps (Figs. 3c,d) reveal marked compositional inhomogeneity, with Sc preferentially concentrated within triangular nanodomains and Al showing a complementary spatial distribution. By superimposing the triangular boundaries extracted from the elemental maps onto the TEM images (Figs. 3e,f), a remarkable spatial correspondence is observed between the Sc-rich regions and the triangular diffraction-contrast features. This correlation strongly suggests that the periodic fringes originate from local lattice distortions associated with Sc segregation. Consistent with this interpretation, previous studies have shown that elevating the Sc fraction within wurtzite ScAlN substantially alters the in-plane lattice parameter and local metal–nitrogen bonding geometry.[42]

Quantitative EDS analysis (Fig. S4, Supporting Information) reveals an exceptionally high local Sc concentration exceeding 55% within the triangular domains, compared with approximately 16% in the surrounding alloy matrix, confirming pronounced nanoscale chemical segregation. Despite such substantial compositional variations, the electron diffraction patterns (Fig. S3, Supporting Information) remain consistent with the predominant wurtzite crystallographic framework, without clearly resolved additional reflections attributable to a secondary crystalline phase. This observation suggests that the Sc-rich nanodomains retain a coherent crystallographic relationship with the surrounding wurtzite matrix, although local structural distortions or nanoscale phase coexistence cannot be completely excluded. Recent work by Hart et al. revealed nanoscale Sc clustering within the wurtzite lattice through cross-sectional atomic-resolution characterization, establishing a connection between local compositional inhomogeneity, lattice distortion, and dielectric response.[43] Our plan-view STEM investigations along the [0001] zone axis further reveal that Sc segregation does not merely manifest as random compositional fluctuations but develops into crystallographically oriented triangular nanodomains spatially correlated with dislocation-associated strain fields. By linking these microstructural features to temperature-dependent electron transport properties, our findings identify the spatial organization of dislocation-associated alloy disorder as a key microscopic factor underlying the pronounced alloy scattering in ScAlN/GaN heterostructures. The apparent absence of pronounced Sc inhomogeneity in cross-sectional STEM-EDS maps (Figs. 1d–f) arises from differing observation geometries. Cross-sectional analysis probes the thin alloy barrier along the growth direction and averages compositional signals over the full projected specimen thickness, whereas plan-view characterization directly resolves the lateral distribution of Sc-rich nanodomains. These triangular domains have lateral sizes of several tens of nanometers, with a characteristic side-length distribution peaking around 40 nm, as quantified in Fig. 4c. Since their lateral extent is much smaller than the projected specimen thickness, the Sc-rich nanodomains contribute only a fraction of the integrated EDS signal and thus yield weak compositional contrast in cross-sectional measurements. as quantified in Fig. 1c. Statistical analysis from plan-view micrographs gives an areal density of approximately $3.3\times10^{10}$ $cm^{-2}$, demonstrating that these Sc-rich nanodomains form spatially correlated assemblies, rather than discrete, random compositional fluctuations.

To further probe the linkage between these Sc-rich nanodomains and crystallographic defects, complementary two-beam diffraction-contrast TEM imaging and elemental mapping were performed (Figs. 3g–l). The dark- and bright-field TEM images (Figs. 3g,h) reveal triangular contrast diffraction features accompanied by prominent strain-sensitive fringes and extended dislocation-line contrast, with contrast evolving markedly upon specimen tilting. The corresponding Al and Sc elemental maps (Figs. 3i,j) reproduce the triangular compositional patterns observed in our prior plan-view STEM measurements, providing independent corroboration of these nanoscale chemical inhomogeneities. By superimposing the Sc-rich domain boundaries extracted from the elemental maps onto the TEM micrographs (Figs. 3k,l), a remarkable spatial correspondence is observed between the compositional nanodomains and the local diffraction-contrast variations. More importantly, threading dislocations reside in proximity to the triangular vertices and edges, suggesting that the formation and spatial arrangement of these nanodomains are closely associated with the dislocation network. This observation is consistent with the intense strain field surrounding dislocations, which produce anisotropic lattice distortions and locally modify the formation energy of high-Sc-content wurtzite ScAlN. In particular, the tensile and compressive strain components surrounding edge dislocations may alter the relative adatom mobilities of Sc and Al atoms, thereby driving spatially nonuniform Sc incorporation during epitaxial growth.

Further insight into the crystallographic origin of the triangular morphology is obtained from the magnified atomic-resolution STEM images in Figs. 3q–t. These images reveal localized disturbance of the atomic-column arrangement by edge dislocations, whose Burgers-vector directions are resolved in the images. The three sets of triangular domain edges run along $\langle 11\bar{2}0\rangle$ directions, corresponding to $\{1\bar{1}00\}$ prism m-planes, analogous to the crystallographically bounded Ga-rich quasi-hexagonal domains reported previously for AlGaN alloys. [44] The resulting triangular morphology suggests that domain evolution is governed not only by elastic interactions but also by anisotropic atomic attachment and step propagation during epitaxial growth. Previous studies have demonstrated that inequivalent A- and B-type steps on wurtzite GaN and AlN (0001) surfaces feature distinct adatom attachment kinetics, favoring the nucleation of triangular islands bounded by crystallographically preferred edges.[45-47] Further analysis of the plan-view Sc maps reveals a preferential spatial partitioning of oppositely oriented triangular domains, consistent with the crystallographic selectivity expected from alternating terrace–step configurations (Fig. S4, Supporting Information). Dislocation-localized strain fields govern Sc segregation by modifying the formation energy and adatom mobility of Sc and Al, whereas terrace-dependent adatom incorporation kinetics dictate the triangular morphology and crystallographic orientation of the resulting Sc-rich domains. In particular, 60° rotation of the step edge swaps A- and B-type step character, yielding anisotropic growth velocities along different crystallographic directions. This anisotropic growth behavior, together with preferential Sc incorporation at strained sites, enables the spontaneous formation of crystallographically oriented triangular Sc-rich nanodomains. Collectively, these results show that the interplay between strain-driven Sc segregation and step-mediated anisotropic growth gives rise to the oriented, spatially correlated Sc-rich nanodomains.

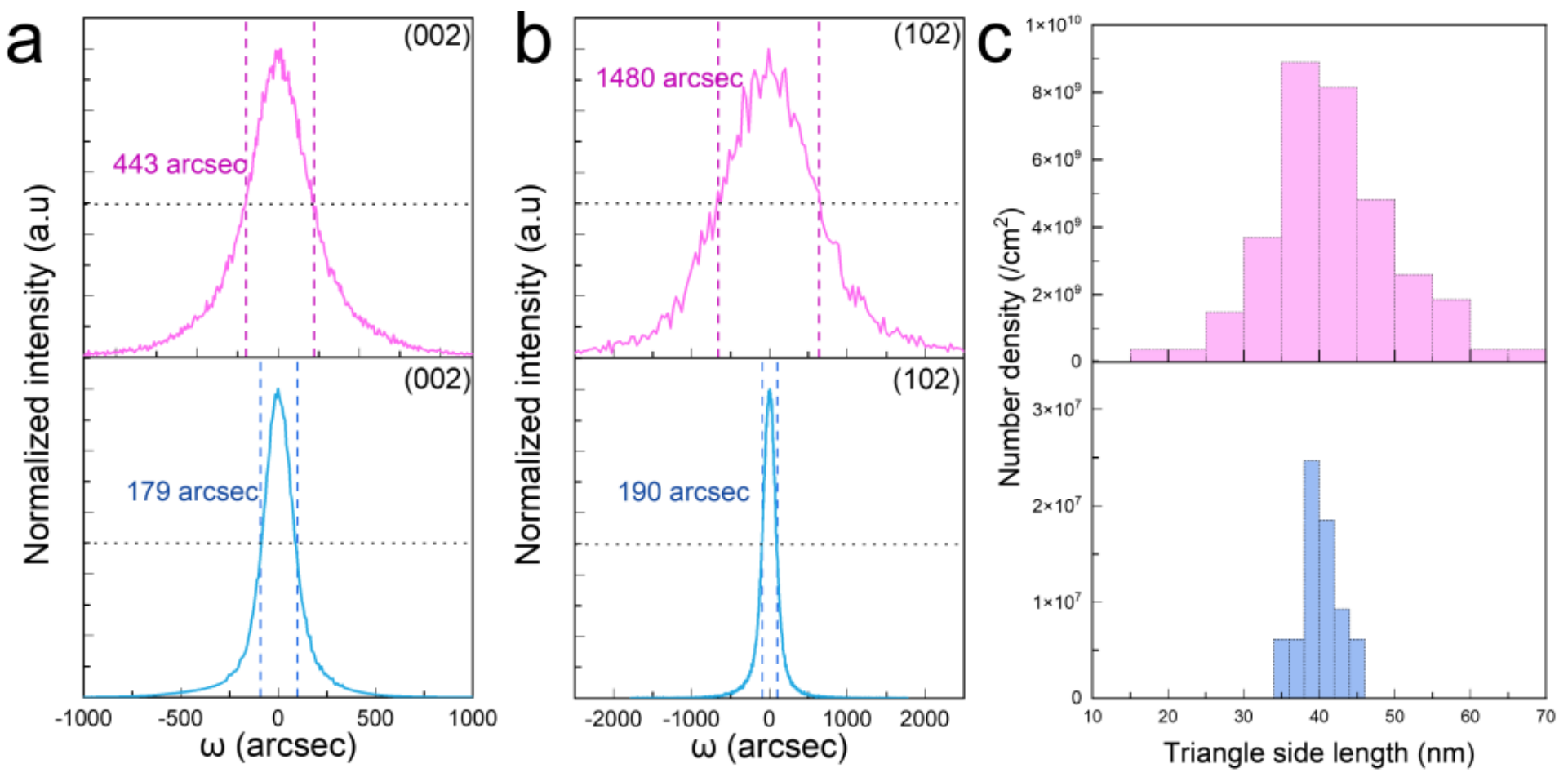

**Figure 4. Correlation between HRXRD rocking-curve broadening and the areal density of triangular Sc-rich nanodomains.** (a,b) HRXRD ω-rocking curves of Samples A (magenta) and B (blue) acquired at the symmetric (002) and asymmetric (102) reflections, with corresponding FWHMs indicated. (c) Side-length histograms and areal number densities of triangular Sc-rich nanodomains in Samples A (top) and B (bottom), extracted from plan-view elemental mapping.

To further validate the linkage between threading dislocations and the formation of triangular Sc-rich nanodomains, we prepared an identical ScAlN/GaN heterostructure on a low-TDD GaN template (denoted Sample B) and systematically compared the areal number density of triangular Sc-rich nanodomains against Sample A using HRXRD and quantitative STEM (Fig. 4). The (002) and (102) XRD rocking-curve FWHMs in Figs. 4a,b, were used to calculate edge- and screw-type TDDs, [48] yielding values of $4.36\times10^{8}$ to $4.58\times10^{7}$ cm$^{-2}$ for Sample B, substantially lower than those of Sample A ($2.78\times10^{10}$ and $4.67\times10^{8}$ cm$^{-2}$), as summarized in Table II. Statistical analysis of STEM elemental maps shows that the characteristic side lengths of triangular domains peak at tens of nanometers for both samples, whereas the areal number density of domains drops from approximately $3.30\times10^{10}$ cm$^{-2}$ in Sample A to $2.84\times10^{8}$ cm$^{-2}$ in Sample B (Fig. 4c). This trend is consistent with the TDD reduction, which is cross-verified by independent STEM dislocation counting (Table II). Specifically, screw-type dislocations terminate atomic steps, and edge dislocations act as step pinning sites,[44] with both disrupting the local terrace–step configuration and thereby modulating the orientation and evolution of Sc-rich nanodomains. In regions of high dislocation density and close dislocation–dislocation interaction, overlapping strain fields and severely perturbed step propagation enable the coexistence of inequivalent A- and B-type step pairs, which gives rise to mixed triangular orientations and a broader domain-size distribution. In contrast, isolated dislocations in the low-TDD GaN template introduce spatially localized strain perturbations. Under this condition, the strain field from each individual dislocation dominates the local incorporation behavior on a given terrace, producing triangular Sc-rich domains with well-conserved orientation and tighter size dispersion. These combined findings reveal that the threading dislocation population — including both its overall density and spatial distribution — governs the spatial ordering, size statistics, and orientational diversity of Sc-rich nanodomains.

**Table II.** Threading dislocation densities (TDDs) determined by HRXRD and direct STEM counting, together with the areal number densities of triangular Sc-rich nanodomains for Samples A and B.

| Sample | **TDD by HRXRD ($cm^{-2}$)** | | | **TDD by STEM ($cm^{-2}$)** | **Areal number density of Sc-rich domains ($cm^{-2}$)** |
|---|---|---|---|---|---|
| | Screw-type | Edge-type | Total | | |
| A | $4.67\times10^{8}$ | $2.78\times10^{10}$ | $2.83\times10^{10}$ | $4.67\times10^{10}$ | $3.30\times10^{10}$ |
| B | $4.58\times10^{7}$ | $4.36\times10^{8}$ | $4.41\times10^{8}$ | $2.31\times10^{8}$ | $2.84\times10^{8}$ |

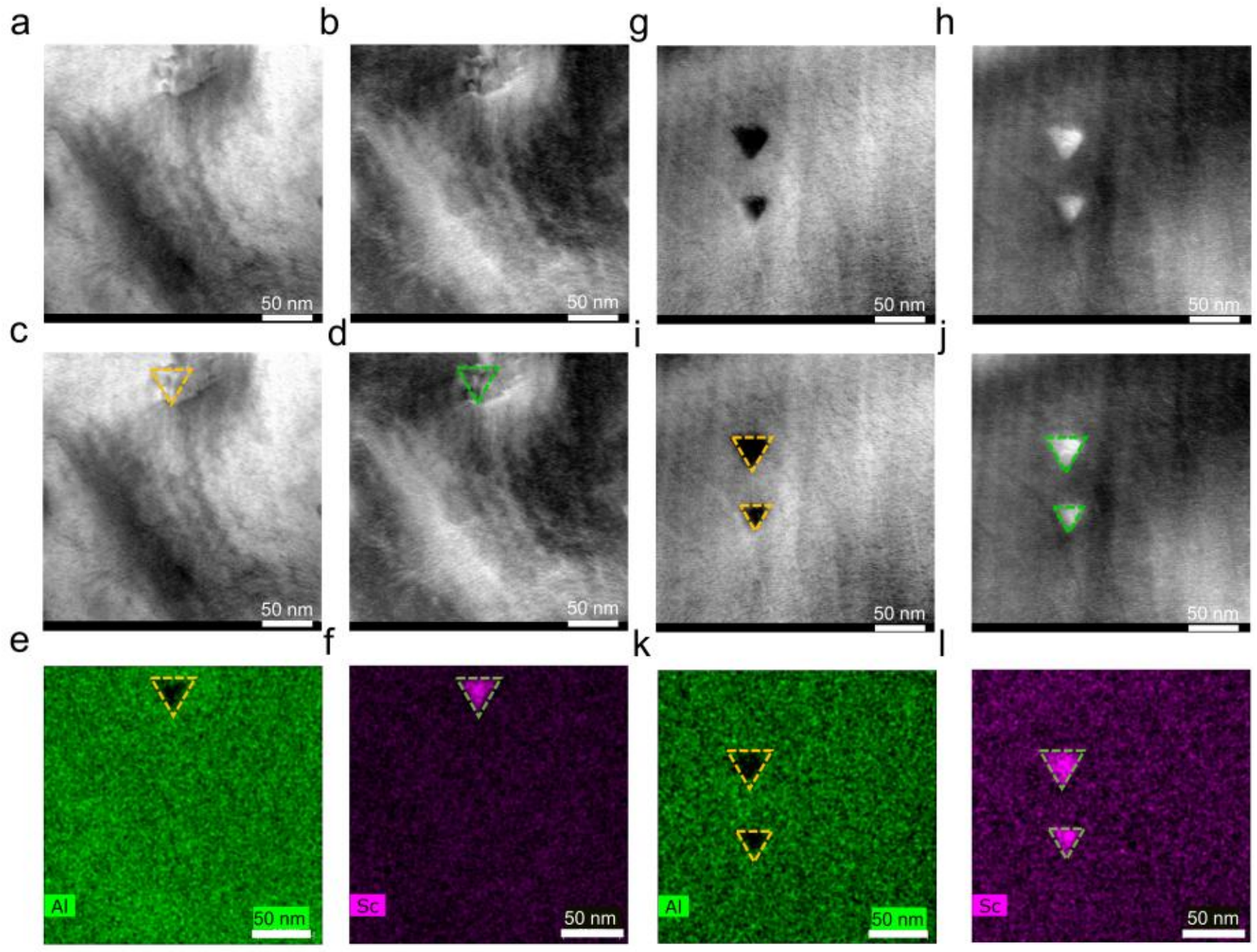

**Figure 5. (a) Comparative microstructural characterization of Sample B.** (a,b) Dark- and bright-field STEM micrograhs, and (c,d) the same micrographs with sparse residual triangular features highlighted. (e,f) Associated Al and Sc elemental maps confirming localized compositional variations. (g,h) Dark- and bright-field TEM micrographs acquired under two-beam diffraction conditions. (i,j) Corresponding annotated micrographs marking the triangular features. (k,l) Al and Sc elemental maps co-registered with the two-beam TEM field of view.

Dark- and bright-field STEM images (Figs. 5a,b), together with annotated features (Figs. 5c,d), directly illustrate the far lower abundance of triangular Sc-rich nanodomains in Sample B. Across four to five matched-magnification fields of view, only one triangular domain was found, and its Sc segregation was verified by EDS mapping (Figs. 5e,f). Notably, these residual domains retain lateral dimensions and local Sc compositional contrast comparable to those observed in Sample A, demonstrating that lowering TDD primarily suppresses domain nucleation without modifying the magnitude of local Sc segregation. Two-beam TEM (Figs. 5g-l) reproduces the strain diffraction contrast seen in Fig. 3 and confirms spatial alignment between compositional domains and strain features. Inspection of dozens of nanodomains shows that neighboring nanodomains tend to adopt a common orientation, consistent with the smoother terrace–step growth morphology expected for low-TDD templates.

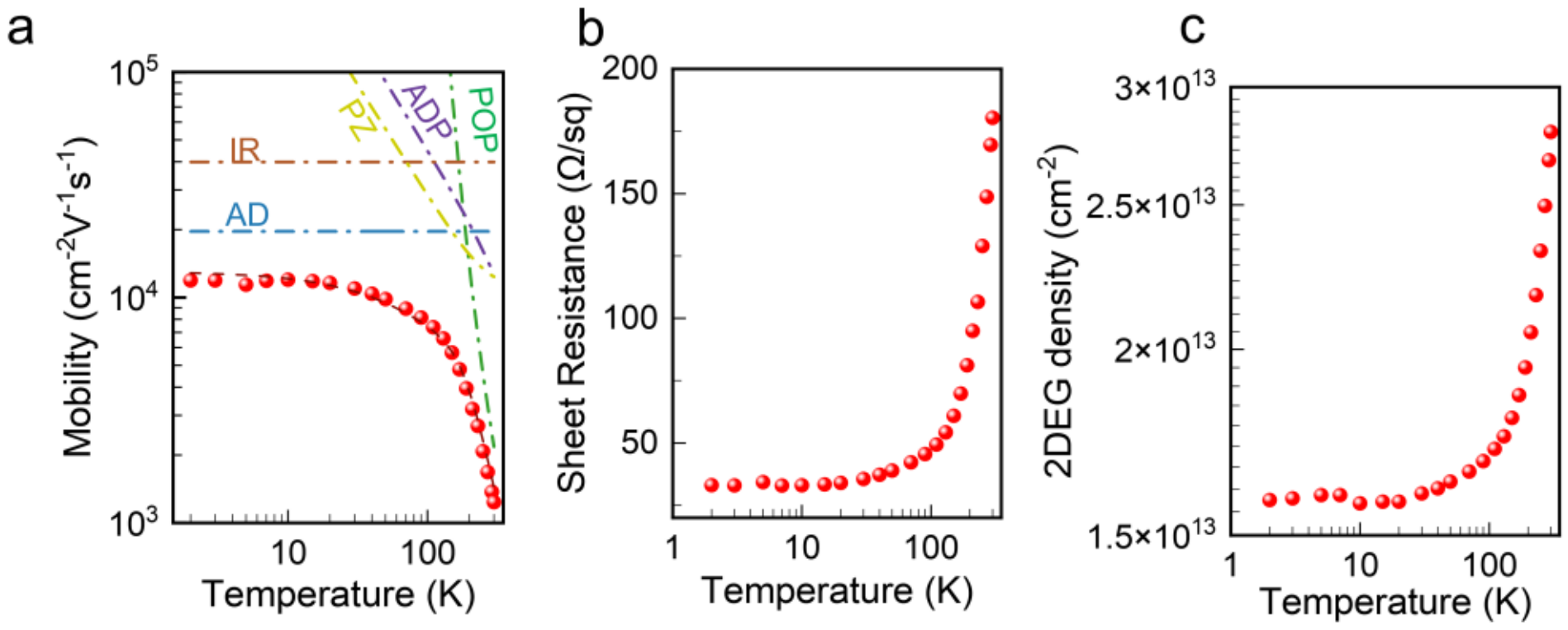

**Figure 6. Temperature-dependent electron transport and scattering analysis of Sample B.** (a) Experimental Hall mobility and calculated mobility contributions from AD, IR, DIS, ADP, PZ, and POP. The total mobility is calculated using Matthiessen's rule. Temperature-dependent (b) sheet resistance and (c) sheet carrier density.

Temperature-dependent Hall measurements of Sample B (Fig. 6) yield a mobility plateau of 11,995 $cm^2 \cdot V^{-1} \cdot s^{-1}$, roughly 6.7 times the value measured for Sample A, corresponding to a sheet resistance of 33.0 $\Omega \cdot \square^{-1}$. This pronounced low-temperature mobility gain indicates reduced extrinsic scattering in Sample B. Using the same transport framework as for Sample A, we have attained a drastically reduced effective alloy scattering potential of 10 eV for Sample B, consistent with suppressed alloy scattering. The coordinated reduction in TDD, triangular Sc-rich nanodomain population, and alloy scattering potential establishes this microstructural mechanism underlying the mobility bottleneck in ScAlN/GaN heterostructures. At 298 K, Sample

B exhibits an apparent Hall mobility of 1290 $cm^2 \cdot V^{-1} \cdot s^{-1}$ and a sheet carrier density of $2.75 \times 10^{13}$ $cm^{-2}$. The apparent carrier density decreases to $1.60 \times 10^{13}$ $cm^{-2}$ upon cooling (Fig. 6c), an indicator of an additional parallel-conduction contribution from the underlying 4-μm GaN buffer template. Using a two-channel conduction model and a realistic mobility range of 200-500 $cm^2 \cdot V^{-1} \cdot s^{-1}$ for electrons in the buffer, the room-temperature 2DEG mobility is projected to be 1860-2070 $cm^2 \cdot V^{-1} \cdot s^{-1}$. Figure 7 benchmarks the low-temperature and room-temperature transport metrics of Sample B against prior ScAlN/GaN heterostructure reports. Together, these results demonstrate that suppressing Sc-rich nanodomain formation by reducing TDD provides a promising strategy for achieving ScAlN/GaN heterostructures with simultaneous high electron mobility and high 2DEG sheet density.

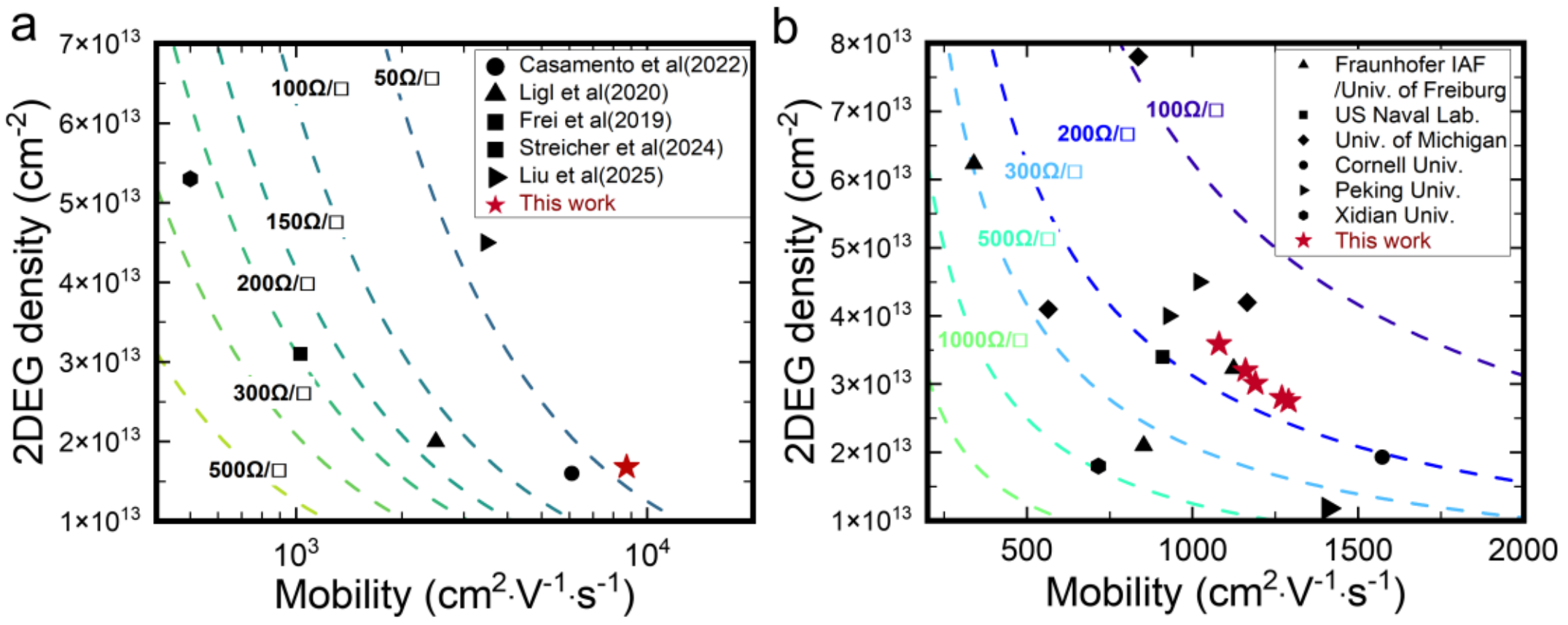

**Figure 7. Benchmark of 2DEG transport performance for Sc-containing nitride/GaN heterostructures.** Comparison of electron mobility versus sheet carrier density at (a) 300 K and (b) 77 K against representative literature reports. Dashed contours denote constant sheet resistance, and red stars mark the data from this work.[8, 18, 19, 26, 29, 49-56]

## Conclusion

In summary, we have investigated the microscopic origin of pronounced alloy scattering in ScAlN/GaN heterostructures by correlating temperature-dependent electron transport properties with threading dislocations, local Sc compositional inhomogeneity, and strain fields. Plan-view STEM and elemental mapping identify orientation-preferred triangular Sc-rich nanodomains spatially coupled to threading dislocations, which we corroborate via diffraction-contrast TEM and atomic-resolution strain mapping. These observations indicate that such nanodomains emerge from the interplay of anisotropic epitaxial incorporation kinetics and local elastic interactions, giving rise to spatially correlated alloy disorder, in contrast to the stochastic, unstructured compositional fluctuations assumed in conventional alloy scattering models, which typically yield low alloy scattering potentials for AlGaN and InGaN. The coordinated reduction in TDD, triangular Sc-rich nanodomain population, and effective alloy scattering potential, together with a substantial rise in low-temperature Hall mobility from 1800 to 11,995 $cm^2 \cdot V^{-1} \cdot s^{-1}$, establishes a direct quantitative structural–transport correlation governing the mobility bottleneck of ScAlN/GaN heterostructures. Collectively, these findings demonstrate that suppressing dislocation-mediated Sc-rich nanodomain formation constitutes a powerful strategy to circumvent alloy-scattering-limited transport, enabling ScAlN/GaN heterostructures with simultaneously high 2DEG mobility and sheet carrier density. This strategy unlocks new performance boundaries for ScAlN/GaN heterostructures and creates tangible opportunities for next-generation high-frequency and high-power electronic technologies.

**Additional information**

Supplementary information is available online.

**ACKOWLEDGMENTS**

The authors are also grateful for the technical support from Nano Fabrication Facility, Platform for Characterization and Test and Nano-X from SINANO, CAS. This study is supported by the National Natural Science Foundation of China (Grant No. 62674233, 62274173). The authors also acknowledge the support from the Open Research Fund of the State Key Laboratory of Widegap Semiconductor Optoelectronic Materials and Technologies (Grant No. SOMAT-2025-05).

**AUTHOR DECLARATIONS**

**Conflict of Interest**

The authors have no conflicts of interest to disclose.

**Author Contributions**

S. Fan conceived the research and designed the experiment. S. Fan supervised the overall research activity. S. Fan and Y. Yin conducted the MBE growth, Hall effect measurement, XRD, STEM/TEM experiment and analysis. S. Fan and Y. Yin wrote the manuscript with comments and revisions from all other authors.

## DATA AVAILABILITY

The data that supports the findings of this study is available within the article and its supplementary information. Additional data is available from the corresponding author upon reasonable request.